\documentstyle[11pt,newpasp,twoside,epsfig]{article}

\begin{document}

\title{Searching For Planetary Transits In The Open Cluster NGC~7789}

\author{D.M. Bramich, K.D. Horne}
\affil{School of Physics \& Astronomy,
       Uni. of St.~Andrews,
       KY16 9SS, UK}

\author{I.A. Bond}
\affil{Institute for Astronomy, 
       Uni. of Edinburgh,
       EH9 3HJ, UK}

\begin{abstract} 
Open clusters are ideal targets for searching for transiting 
Hot Jupiters. They provide a relatively large concentration
of stars on the sky and cluster members have similar metallicities,
ages and distances. Fainter cluster members are likely to show deeper
transit signatures, helping to offset sky noise contributions. A 
survey of open clusters will then help to characterise the Hot
Jupiter fraction of main sequence stars, and how this may depend
on primordial metallicity and stellar age.

We present results from 11 nights of observations of the open cluster 
NGC~7789 with the WFC camera on the INT telescope in La Palma. From
684 epochs, we obtained lightcurves and $\bv$ colours for $\sim$25600 stars, 
with $\sim$2400 stars with better than 1\% precision. We expect to detect
$\sim$1 transiting Hot Jupiter in our sample assuming that 1\% of stars host a
Hot Jupiter companion.
\end{abstract} 

\section{Introduction}

The surprising existence of short period ($\sim$4 days) Jupiter mass
extra-solar planets (termed ``Hot Jupiters'') confirmed by
radial velocity (RV) measurements in the last 8 years has shown
that our own solar system is certainly not typical. The  
class of Hot Jupiter planets ($P \leq 10$ days and $M \sin{i} \leq10 M_{J}$)
makes up $\sim$15\% (17 out of 117 as of 01/10/03) of the planets discovered by the RV technique to date 
(Schneider 1996) and $\sim$1\% of nearby solar type stars host 
such a companion (Butler et al. 2000). To date there are only two confirmed transiting extra-solar planets,
HD 209458b (Charbonneau et al. 2000; Brown et al. 2001) discovered first by the RV method, and OGLE-TR-56 
discovered first by the transit method (Udalski et al. 2002; Konacki et al. 2003).
Using geometric considerations and the planet/star properties
of HD 209458b as a typical Hot Jupiter system, we get the probability of a full transit given that a 
solar type star has a Hot Jupiter as $\sim$11\%. This fits in nicely with the discovery of 17 Hot 
Jupiters to date via RV but only one transiting Hot Jupiter in this sample. 
More logically one should observe large numbers of stars in parallel to obtain transit candidates
and then use RV to follow up.
Recently we are starting to see the fruits of current transit surveys. OGLE have produced of the order 
of 100 transit candidates over two seasons (Udalski et al. 2002a,b; Udalski et al. 2003)
and EXPLORE have produced a possible 4 transiting planets (Yee et al. 2003), all of which have yet to be 
confirmed by detailed spectroscopic follow up.

The study of open clusters for transiting planets has a number of advantages over fields
in other parts of the sky or galactic plane. While providing a relatively large
concentration of stars on the sky (but not so large as to cause blending problems
as in the case of globular clusters observed from the ground), they also provide
a set of common stellar parameters for the cluster members. These are metallicity,
age, stellar crowding and radiation density. All cluster members lie at roughly the same
distance aswell allowing magnitude and colour to be directly related to star radius/mass
for main sequence cluster members. A trend that host stars tend to be metal rich (Santos et al. 2003) is 
one of the results that will be confirmed or refuted from the determination of the fraction of stars 
hosting a Hot Jupiter (from now on referred to as the Hot Jupiter fraction) in open clusters.
%
%The observations of open cluster NGC~7789 is the subject of this paper. The main parameters of the 
%cluster are shown in Table 1. For a good review of previous relevant work on this cluster see Gim et al. 
%(1998). Section 2 reports the observations made, section 3 presents in detail the data 
%reduction/photometry, section 4 presents the astrometry and colour data, section 5 reports on the 
%transit detection algorithm and makes an estimate of the likely planet catch of the data set, 
%and section 6 outlines our conclusions and future work.
%
\begin{table}
\begin{center}
\caption{Properties of the open cluster NGC~7789. Data taken from 
         http://obswww.unige.ch/webda by Mermilliod, J.C. and the SIMBAD database.}
\bigskip
\begin{tabular}{|c|c|c|c|c|c|}
\hline
RA (J2000.0)  & 23$^{h}$ 57$^{m}$ & $b$           & -5\fdg37        & Age (Gyr)   & 1.7   \\ 
\hline
Dec (J2000.0) & +56\deg 43\arcmin & Distance (pc) & 1900            & Metallicity & -0.24 \\
\hline
$l$           & 115\fdg48         & Radius        & $\sim$16\arcmin & $E(\bv)$    & 0.22  \\
\hline
\end{tabular}
\end{center}
\end{table}

\section{Observations}

We observed the open cluster NGC~7789 (see Table 1) using the 2.5m Isaac Newton Telescope (INT) on La 
Palma in the Canary Islands during the 11 nights from 10th to 20th August 2000. We used the Wide Field 
Camera (WFC) which consists of a 4 EEV CCD mosaic where each CCD is 2048x4096 pixels. The pixel 
scale is 0.33\arcsec/pix and field of view $\sim$0.5\deg x 0.5\deg. The gain and readout noise values 
for each chip were calculated automatically during the preprocessing stage of the data reduction (Section 3).
%(see Section 3.1). 
The mosaic field was centred on NGC~7789 at $\alpha=23^{h}57^{m}30\fs8$ and 
$\delta=+56\deg43\arcmin41\farcs9$. The procedure for each night was to obtain $\sim$5 bias frames 
and $\sim$8 sky flat frames at both the beginning and end of the night. Observations on NGC~7789 
consisted of ten 300s exposures followed by a single bias frame repeated throughout the whole night. 
With a readout time of 100s and various losses due to bad weather/seeing and telescope jumps, this 
resulted in a total of 684x300s exposures in Sloan $r^{\prime}$. We also took 5 images of NGC~7789 with 
varying exposure times in Harris $V$, along with 5 sky flat frames, in order to 
provide us with the necessary colour information.  

\section{CCD Reductions And Photometry}

Each chip was treated independently for the purpose of the reductions: bias pattern removal, overscan 
correction, flat fielding etc. The reduction process was carried out by a single C-shell/IRAF script that 
runs according to a user defined parameter file. Bad pixels were flagged in a user defined detector bad 
pixel mask, and ignored where relevant.
%
%The photometry on the reduced science frames in the Sloan $r^{\prime}$ filter was done using the method 
%of difference image analysis (Alard \& Lupton 1998; Alard 2000). Our implementation of this procedure was
%adapted from the code written for the MOA project (Bond et al. 2001), and it consists of three automated 
%scripts. Bad pixels are propagated through the scripts in the correct fashion.

The photometry on the reduced science frames in the Sloan $r^{\prime}$ filter was done using the method
of difference image analysis (Alard \& Lupton 1998; Alard 2000). Our implementation of this procedure was
adapted from the code written for the MOA project (Bond et al. 2001), and it consists of three automated  
scripts. Bad pixels are propagated through the scripts in the correct fashion. The first script was 
used to construct a mean reference frame from a set of 13 consecutive best seeing ($\sim$ 1\arcsec) frames, 
and produce an associated star list with magnitudes measured by DAOPhot PSF fitting.
The second script produced the difference images (see Bond et al. 2001) and the third script measured the 
differential flux on each difference image via optimal PSF scaling at the position of each star.

A lightcurve for each star was constructed by the addition of the differential fluxes to the star magnitude 
as measured on the reference frame. 
%Uncertainties are propagated in the correct analytical fashion. 
Flux measurements were rejected for a $\chi^{2} pix^{-1} \geq 5.0$ for the PSF scaling, and for PSFs with a 
FWHM$\geq$7.0 pix, in order to remove bad measurements. Hence all the stars have varying numbers of epochs 
measured. Lightcurves with less than 300 epochs (out of a possible 684) were rejected. This analysis 
produced 8631 lightcurves on chip 1, 7625 lightcurves on chip 2, 8411 lightcurves on chip 3 and 8830 
lightcurves on chip 4 (centred on the cluster). Figure 1 shows a diagram of the scatter in the 
lightcurves against magnitude for chip 2.

\begin{figure}[t]
\begin{center}
\epsfig{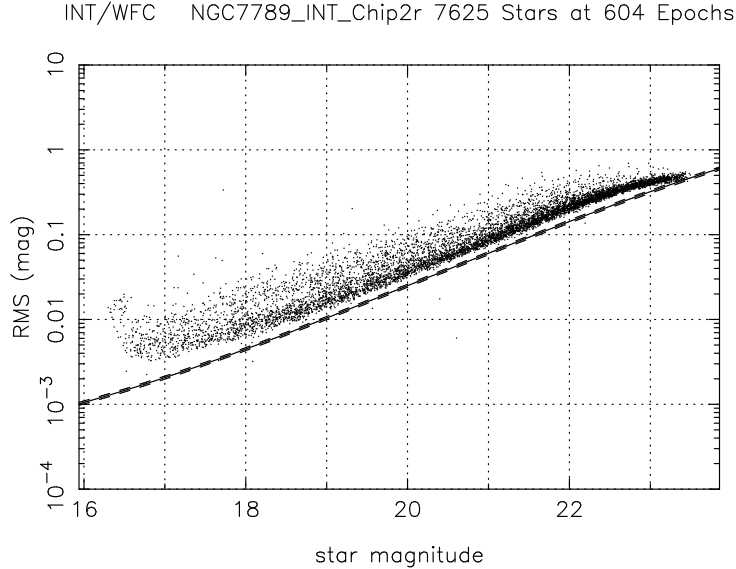}
\caption{A plot of standard deviation of the lightcurves against mean instrumental
         Sloan $r^{\prime}$ magnitude for all stars from chip 2. The lower curve represents the 
         theoretical noise limit for photon and readout noise.}
\end{center}
\end{figure}

\section{Astrometry And Colour Data}

The astrometry was done by matching 358 stars from the four reference frames (one for each chip) with 
%the USNO-B1.0 star catalogue (Monet et al. 2003) using a field overlay in the image display tool 
the USNO-B1.0 star catalogue using a field overlay in the image display tool
%GAIA (Draper 2000). The WFC suffers from pincushion distortion, hence it was necessary to fit a 
GAIA. The WFC suffers from pincushion distortion, hence it was necessary to fit a
9 parameter astrometric solution to the reference frames in order to obtain sufficiently
accurate equatorial coordinates for all the stars. The 9 parameters are made up of 6 parameters to 
define the linear transformation between pixel coordinates and equatorial coordinates, 2 parameters to 
define the plate centre and 1 parameter to define the radial distortion coefficient. The 
%starlink package ASTROM (Wallace 1998) was used to do the fit and the achieved accuracy was $\sim$0.4 
starlink package ASTROM was used to do the fit and the achieved accuracy was $\sim$0.4
arcsec RMS radially for the 358 matching stars.

The two best images in the Harris $V$ filter were aligned with the Sloan $r^{\prime}$ reference frame 
for each chip and the magnitudes of the stars were measured using DAOPhot PSF fitting.
%in the same way as they were measured on the reference frame in section 3. 
Magnitude measurements were averaged and any 
stars without two measurements were treated as having an unknown Harris $V$ magnitude.
This was done so that airmass differences between the two Harris $V$ images did not have to be taken 
into account. In order to calibrate the instrumental Harris $V$ and Sloan $r^{\prime}$ magnitudes, 
Johnson $BV$ data kindly supplied by B. Mochejska was used. For a full description of this data set,
see Mochejska \& Kaluzny (1999). Stars were identified as matches between the Johnson $BV$ data set and 
the Harris-Sloan $Vr^{\prime}$ data set if they were within 0.5\arcsec of each other (using the equatorial 
coordinates from each data set) and if there were no other stars from the Harris-Sloan $Vr^{\prime}$ data set 
within 4.0\arcsec. This resulted in 286 matches for chip 1, 100 matches for chip 2, 62 matches for chip 3 and 
812 matches for chip 4. A linear transformation of the form:
\begin{equation}
\left(\begin{array}{c} V \\ \bv \end{array}\right)
= 
\left(\begin{array}{cc} 1 & m_{1} \\ 0 & m_{2} \end{array}\right) 
\left(\begin{array}{c} r^{\prime}_{ins} \\ r^{\prime}_{ins} - v_{ins} \end{array}\right)
+ 
\left(\begin{array}{c} c_{1} \\ c_{2} \end{array}\right)
\end{equation}
was solved for by minimising the chi squared of the fit for the matching stars on each chip. The symbols $B$ and 
$V$ represent the Johnson $BV$ magnitudes, and the symbols $r^{\prime}_{ins}$ and $v_{ins}$ represent the 
Sloan $r^{\prime}$ and Harris $V$ instrumental magnitudes respectively. 
%The results 
%of the fits for these transformations are listed in Table 2. 
Figure 2 shows an instrumental colour 
magnitude diagram for chip 4, the chip containing the cluster, where the cluster main sequence can be 
clearly seen. Around 76\% stars in our data set have calibrated colour measurements.

\begin{figure}[t]
\begin{center}
\epsfig{file=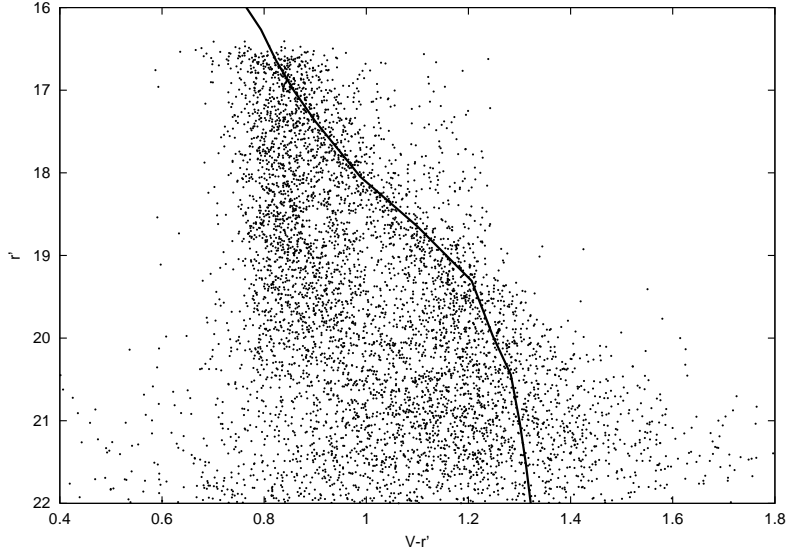,angle=270,width=0.8\linewidth}
\caption{Instrumental colour-magnitude diagram for $r^{\prime}$ and $V-r^{\prime}$. The solid line
         represents the theoretical main sequence for cluster stars at a distance of 1900pc and 
         with $E(B-V)=0.22$.}
\end{center}
\end{figure}
% 
% \begin{table}
% \begin{center}
% \caption{The coefficients of the linear transformation between instrumental Harris $V$
%         and Sloan $r^{\prime}$ magnitudes and Johnson $BV$ magnitudes.}
% \bigskip
% \begin{tabular}{|c|c|c|c|c|}
% \hline
% Chip No. & $m_{1}$ & $c_{1}$ & $m_{2}$ & $c_{2}$ \\
% \hline
% 1 & 1.006 & -0.496 & 1.919 & -0.851 \\
% 2 & 0.915 & -0.225 & 1.741 & -0.513 \\
% 3 & 1.149 & -0.277 & 2.101 & -0.659 \\
% 4 & 0.871 & -0.160 & 1.697 & -0.497 \\
% \hline
% \end{tabular}
% \end{center}
% \end{table}

\section{Transit Detection}

A modified matched filter algorithm was used in order to detect the transit candidates. For each 
lightcurve, a set of 12 transit durations in a geometric sequence starting at 0.5 hours
with a geometric factor of $\sim$1.23, and ending at 5.0 hours, was generated. For each transit duration   
$\Delta t$, the lightcurve was stepped through from the start of the lightcurve $T_{start}$ to the end of 
the lightcurve $T_{end}$ in time steps of $\Delta t / 4$. At each time $t_{0}$, a box car transit curve 
of total length $5 \Delta t$, transit duration $\Delta t$ and time of central transit $t_{0}$ was 
fitted via least squares, and a constant magnitude of total length $5 \Delta t$ and central time $t_{0}$ 
was also fitted via least squares. The fits were only carried out if there were at least 3 data points 
during the ``in transit'' and at least 8 data points during the ``out of transit''. For each lightcurve, 
$\Delta t$ and $t_{0}$ a transit statistic was calculated as defined by:
\begin{equation}
TRA_{STAT} = \frac{ \chi^{2}_{tra} - \chi^{2}_{const} }
                  { \left(\frac{ \chi^{2}_{out} }{ N_{out} - 1 }\right) }
\end{equation}
where $\chi^{2}_{tra}$ is the chi squared of the box car transit fit, $\chi^{2}_{const}$ is the chi 
squared of the constant fit, $\chi^{2}_{out}$ is the chi squared of the ``out of transit'' for 
the boxcar transit fit and $N_{out}$ is the number of data points ``out of transit''.
The statistic $TRA_{STAT}$ is effectively the S/N squared of the fitted transit signal renormalised 
to the reduced chi squared of the ``out of transit''. This modified matched filter algorithm was designed 
to help downweight systematic errors with $\chi^{2}_{out} / (N_{out} - 1) > 1$ (and 
serendipitously, variables) since transit signals should have $\chi^{2}_{out} / (N_{out} - 1) \sim 1$. 
An example boxcar transit fit highlighting a transit candidate is shown in Figure 3.

\begin{figure}[t]
\begin{center}
\epsfig{file=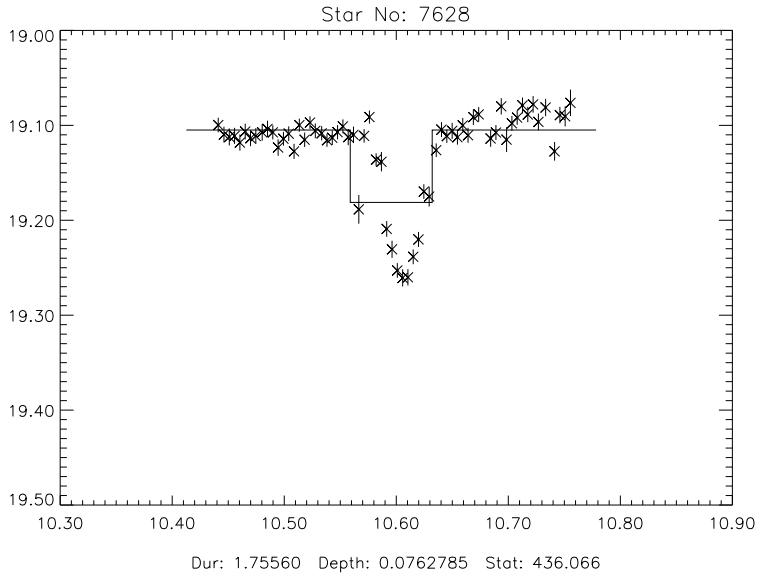,angle=0,width=0.8\linewidth}
\caption{An example box car transit fit showing the ``in transit'' and ``out of transit'' zones. The 
         horizontal axis is time (days) and the vertical axis instrumental Sloan $r^{\prime}$ magnitude.
         ``STAT'' is the value of the transit statistic for this fit.}
\end{center}
\end{figure}

Using the transit detection algorithm, it would be useful to know how many Hot Jupiter planets to expect 
to detect given the data. To do this a Monte Carlo simulation was carried out. For each lightcurve/star 
with a colour, a set of 1000 Hot Jupiters was generated with radii of 1.40$R_{J}$, $\log P$ 
uniformly distributed between $\log(3.0$ days$)$ and $\log(30.0$ days$)$, $\cos i$ uniformly 
distributed between 0 and 1, and $t_{0}$ uniformly distributed between $T_{start}$ and 
$T_{start} + P$ ($P$ and $i$ represent orbital period and orbital inclination respectively). The lower 
limit on $\log P$ is taken from the fact that all RV exoplanets have $P \geq 3.0$ days,  
and the upper limit on the period is arbitrary. The transit lightcurve for each Hot Jupiter was added to 
the current lightcurve in turn and the transit statistic calculated at each transit. The fraction of Hot 
Jupiters detected (Hot Jupiter detection probability) was calculated as a function of the transit 
statistic threshold and the number of transits detected. In order to calculate the false alarm 
probability also as a function of the transit statistic threshold and the number of transits detected, 
the transit statistic was calculated at each transit without adding in the transit lightcurve, but using 
the current transit duration. 

Figure 4 shows a plot of the number of expected planets and false alarms for all stars with colours in 
the data. A transit statisitic threshold of 100 (S/N$\sim$10) for one detected transit signature seems 
a good choice in order to limit the number of expected false alarms but maintain a relatively high number 
of expected detections. Such a choice yields 1.93 expected false alarms and 100.7 expected Hot Jupiter 
detections. Assuming that $\sim$1\% of stars host a Hot Jupiter, then we expect $\sim$1 transiting 
planet. We cannot limit our transit candidates to those that show two transit signatures since this 
would mean that we expect to detect no planets due to the poor period coverage of the data set.

\begin{figure}[t]
\begin{center}
\epsfig{file=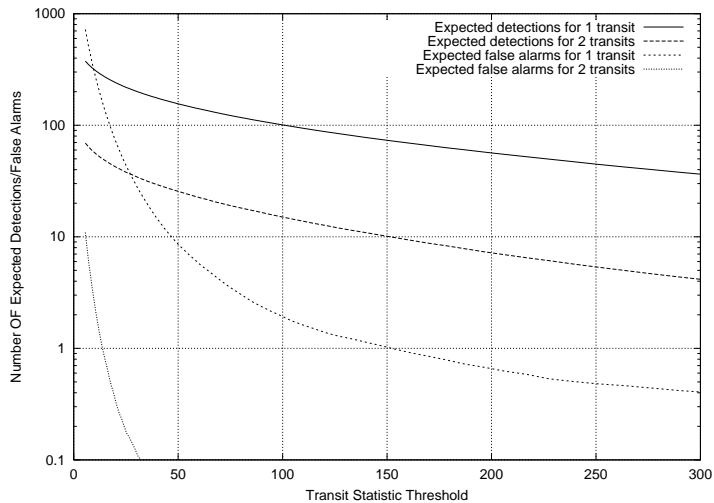,angle=270,width=0.72\linewidth}
\caption{A plot of number of expected detections/false alarms for a 1.40$R_{J}$ Hot Jupiter
         against transit statistic threshold.}
\end{center}
\end{figure}

\section{Conclusions And Future Work}

In the search for our transit candidates we have developed an accurate, efficient and fast photometry 
pipeline that uses the raw data from the telescope to deliver lightcurves. This is important 
considering the high quantity of data that may arise from a transit survey. The pipeline has also been 
applied successfully to other data sets including the PLANET 2002 transit data.

We hope to assign a cluster membership probability to each star using the colour, magnitude and 
position data to construct a 3D probability density function via maximum likelihood fitting. This will 
allow us to split the star sample in a statistical sense into cluster and field stars.

Having applied the transit detection algorithm to our data set we have a number of transit 
candidates (see Figure 3) and variable stars (see Figure 5). 
We may supply an estimate of the planet radius for each candidate since we have star colours (assuming 
the stars to be main sequence), but we will need more data in order to confirm the planetary status. 
For the transit candidate shown in Figure 3, we derive a minimum planetary radius of $0.97^{+0.26}_{-0.17} R_{J}$
by fitting a centrally transiting Hot Jupiter orbiting a linear limb darkened star ($\mu = 0.5$).
Confirmation in the magnitude range of our star sample will take the form of ruling out possible transit 
mimics, like grazing stellar binaries etc. To do this we will need two colour time series observations. 
Radial velocities may constrain the mass of the planet for the brighter candidates. 

When the status of our transit candidates is confirmed we may use the detection probabilities and the 
cluster membership probabilities to estimate the Hot Jupiter fraction for the cluster and the field stars. 
This may then be compared to the current estimates of the Hot Jupiter fraction for main sequence stars. 

\begin{figure}[t]
\begin{center}
\epsfig{file=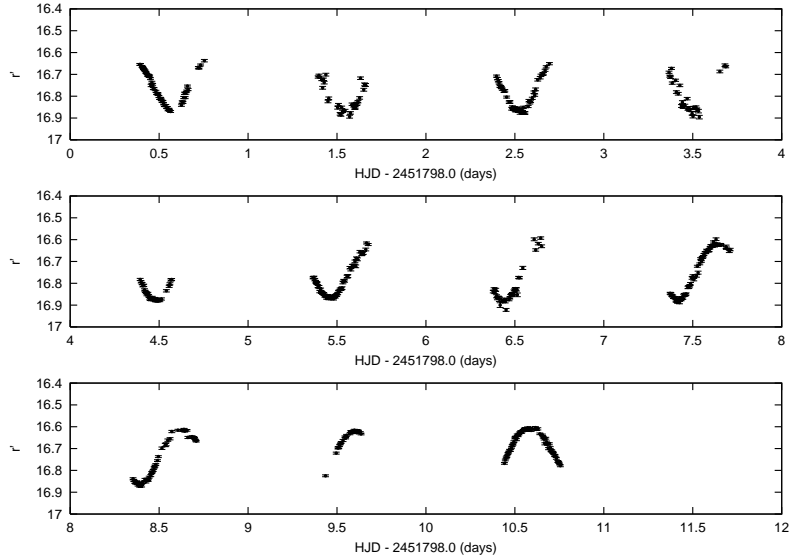,angle=270,width=0.8\linewidth}
\caption{A variable star lightcurve detected by the transit detection algorithm.}
\end{center}
\end{figure}

\end{document}